\documentclass[sigconf]{acmart}

\AtBeginDocument{%
  }

\setcopyright{acmlicensed}
\copyrightyear{2018}
\acmYear{2018}
\acmDOI{XXXXXXX.XXXXXXX}

\acmConference[Conference acronym 'XX]{Make sure to enter the correct
  conference title from your rights confirmation emai}{June 03--05,
  2018}{Woodstock, NY}
\acmISBN{978-1-4503-XXXX-X/18/06}

\usepackage{multirow}
\usepackage{makecell}
\usepackage{fancybox}
\usepackage{listings}

\newenvironment{myresultbox}[1][]{
  \begin{figure}[H]
  \setlength{\fboxsep}{10pt} % Adjust the padding as needed
  \centering
  \begin{Sbox}
  \begin{minipage}{\dimexpr\linewidth-2\fboxsep-2\fboxrule} % Adjust the width as needed
  \textbf{#1} \\
}{
  \end{minipage}
  \end{Sbox}
  \fbox{\TheSbox}
  \end{figure}
}

\begin{document}

\title{HITS: High-coverage LLM-based Unit Test Generation via Method Slicing}

\author{Zejun Wang}
\authornote{Both authors contributed equally to this research.}
\email{zejunwang@pku.edu.cn}
\orcid{0000-0001-7701-998X}
\author{Kaibo Liu}
\authornotemark[1]
\email{liukb@pku.edu.cn}
\affiliation{%
  \institution{Key Lab of HCST (PKU), MOE; SCS}
  \city{Beijing}
  \country{China}
}

\author{Ge Li}
\affiliation{%
  \institution{Key Lab of HCST (PKU), MOE; SCS}
  \city{Beijing}
  \country{China}}
\email{lige@pku.edu.cn}

\author{Zhi Jin}
\affiliation{%
  \institution{Key Lab of HCST (PKU), MOE; SCS}
  \city{Beijing}
  \country{China}
}
\email{zhijin@pku.edu.cn}

\renewcommand{\shortauthors}{Wang et al.}

\begin{abstract}
Large language models (LLMs) have behaved well in generating unit tests for Java projects. However, the performance for covering the complex focal methods within the projects is poor. Complex methods comprise many conditions and loops, requiring the test cases to be various enough to cover all lines and branches. However, existing test generation methods with LLMs provide the whole method-to-test to the LLM without assistance on input analysis. The LLM has difficulty inferring the test inputs to cover all conditions, resulting in missing lines and branches. To tackle the problem, we propose decomposing the focal methods into slices and asking the LLM to generate test cases slice by slice. Our method simplifies the analysis scope, making it easier for the LLM to cover more lines and branches in each slice. We build a dataset comprising complex focal methods collected from the projects used by existing state-of-the-art approaches. Our experiment results show that our method significantly outperforms current test case generation methods with LLMs and the typical SBST method Evosuite regarding both line and branch coverage scores.
\end{abstract}

\begin{CCSXML}
<ccs2012>
   <concept>
       <concept_id>10011007.10011074.10011099.10011102.10011103</concept_id>
       <concept_desc>Software and its engineering~Software testing and debugging</concept_desc>
       <concept_significance>500</concept_significance>
       </concept>
   <concept>
       <concept_id>10010147.10010178.10010179</concept_id>
       <concept_desc>Computing methodologies~Natural language processing</concept_desc>
       <concept_significance>300</concept_significance>
       </concept>
 </ccs2012>
\end{CCSXML}

\ccsdesc[500]{Software and its engineering~Software testing and debugging}
\ccsdesc[300]{Computing methodologies~Natural language processing}

\keywords{Unit Test Generation, Large Language Model, Program Decomposition, Program Slicing, Testing and Analysis, AI for SE}

%%
%% This command processes the author, affiliation and title
\maketitle

%% The Plan:
%% P1: Briefly Introduce automatic unit test generation (summary of Section 2.1)
%%  1.1 What is automatic unit test generation; 1.2 Categorization of classic methods and state-of-the-art methods for each category; 1.3 Trend of using LLM for unit test generation
%% P2: Explain the challenge that testing complex methods is far from satisfaction
%%  2.1: what are complex methods; 2.2 Previous methods' performance DROP on complex methods (compared with the average coverage scores); 2.3 Causes for low coverage scores
%% P3. Propose the solution to the challenge: To generate unit tests for a slice of the focal complex method. (Summary for Section 3)
%%  3.1 What is a lice of the focal complex method; 3.2 Why this work; 3.3 How to generate tests for a slice
%% P4. Brief Introduction to the experiments and results
%%  4.1 The baselines and dataset; 4.2 summarization of the improvements in quantity
\section{Introduction}
High coverage scores are essential metrics for automatic unit test generation methods to pursue \cite{evosuite}. Traditional search-based test generation methods (SBST) incorporate various strategies to lead their search to cover missing branches. For example, Randoop\cite{randoop} uses feedback as the search guidance, and Evosuite \cite{evosuite} utilises the genetic algorithm to explore the input space efficiently. Deep learning methods can generate unit tests with higher coverage scores than SBST methods due to the knowledge learned from human-written tests. For instance, ATHENATEST\cite{athenatest} outperforms Evosuite\cite{evosuite} in the branch coverage on NumberUtils of Lang-1-f by the percentage of 3.5\%. Recent works show that large language models can further enlarge the improvements considering average coverage scores on all methods of the project to test. The researchers of ChatTester \cite{chattester} evaluate ChatGPT, which has improved the statement coverage from Evosuite's 68.0\% to 82.3\% evaluated on their datasets. ChatUniTest\cite{chatunitest} reports that they have improved the average branch coverage from Evosuite's 86.59\% to 90.6\%, and the line coverage score is improved from 80.02\% to 89.36\% on their datasets.
% \kb{The methods mentioned here all surpass Evosuite, yet none of them are used as baselines. On the other hand, the used baselines do not seem to be as good as Evosuit. Reviewers would suspect that your baselines are too weak.}

However, reaching high coverage scores when testing complex methods is still challenging for existing automatic unit test generation methods. The complex methods are the functions with cyclomatic complexities\cite{cyclomatic} greater than 10\cite{sanusi2020development}. According to Tom McCabe, these methods have moderate or higher risk\cite{sanusi2020development}. Thus, reaching high coverage scores for complex methods is essential. Unfortunately, existing unit test generation tools perform poorly on complex methods. 

We evaluate ChatUniTest, a novel published test generation tool based on large language models\footnote{\url{https://github.com/ZJU-ACES-ISE/ChatUniTest}}, as an illustration. ChatUniTest is assessed on two projects, Commons-CLI and Datafaker. The former is included in the large language model's training data, while the latter is excluded. The evaluation results are listed in Table \ref{tab:perform_drop}. `All` means the average coverage scores on all methods within the project-to-test, and `Complex` is for the complex methods only. According to the results, both the line and branch coverage scores of ChatUniTest significantly drop when applied to test complex methods.

\begin{table}[htbp]
    \centering
    \caption{ChatUniTest Coverage Drop on Complex Focal Methods}
    \begin{tabular}{cccc}
    \toprule
      &  & line coverage & branch coverage\\
    \midrule %  
      \multirow{2}{*}{Commons-CLI}   & All & 93.52\% & 96.36\%   \\
            & Complex & 51.17\% (-42.35\%) & 45.83\% (-50.53\%)\\
    \midrule
      \multirow{2}{*}{Datafaker} & All & 86.07\% & 89.86\% \\
        & Complex & 36.75\%	(-49.32\%) & 28.75\% (-61.11\%) \\
    \bottomrule
    \end{tabular}
    % \caption{Coverage Drop on Complex Methods\kb{complex method -> complex focal method}}
    \label{tab:perform_drop}
\end{table}

To address the challenge, we take advantage of the divide-and-conquer algorithm. We use a large language model to generate test cases separately for code slices decomposed from the original method-to-test and combine the test cases as the complete test suite. A complex method is a procedure with a few steps to solve a problem. A code slice comprises successive statements describing a solution step. Generating tests for individual slices helps improve the coverage scores by simplifying the large language model's analysis scope of each generation round. If focusing on a particular code slice, the large language model will analyse fewer branches than considering the whole method-to-test. Besides, the LLM does not have to consider all the concrete execution paths in the previous slices. To reach a branch in the slice-to-cover, only one feasible execution path through the previously analysed slices is enough, making synthesising the desired test inputs easier.

\begin{figure}[htbp]
    \centering
    \includegraphics[width=1\linewidth]{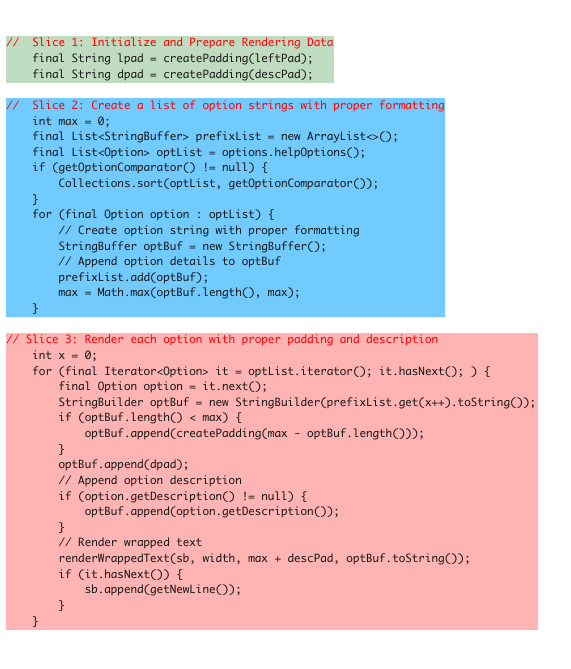}
    \caption{Example of Method Decomposed into Slices}
    \label{fig:slice-example}
    \Description{HelpFormatter.renderOptions() in the project Commons-CLI is divided into three steps by ChatGPT and revised by our human programmers}
\end{figure}

Figure \ref{fig:slice-example} illustrates decomposing a complex method into slices by problem-solution steps. The method `HelpFormatter.renderOptions()' of Commons-CLI is decomposed into three slices by ChatGPT and revised by the senior programmers in our group. This method solves the problem of rendering the specified `Options' and returns the rendered content in a `StringBuffer'. ChatGPT decomposes the method into three slices, each representing a step of the solution and marked with an individual background colour in the figure. By focusing on one slice, the LLM has fewer branches to analyse and synthesising desired test cases becomes easier.  We take the analysis on `optList' as an example. When focusing on Slice 2, the LLM only needs to consider whether `optList' is empty for full line and branch coverage related to `optList'. The concrete element within `optList' does not need to be considered. When focusing on Slice 3, the LLM needs only to analyse one object in `optList': its `length', `getDescription', and `hasNext'. Reducing the analysis scope makes the guidance for high-coverage test cases more informative and precise.

In practice, we propose HITS, which generates high-coverage test suites specifically for complex methods. First, the information on the dependencies of the method-to-test is retrieved via static analysis. Then, we adopt the chain-of-thought technique to instruct the large language model to perform method decomposition and test generation sequentially. We designed plans and examples to let the large language model master method decomposition and generate tests for a specific method via in-context learning. For the tests that failed to execute, we use self-debug\cite{selfdebug} to fix them. We share the relative artefact anonymously\footnote{\url{https://anonymous.4open.science/r/SlicePromptTest4J-6CF1/}}. 

We evaluate HITS on ten open-source projects collected from the Internet used by or related to previous works. HITS is compared with the traditional tool Evosuite, and state-of-the-art LLM-based methods, ChatUniTest \cite{chatunitest}, ChatTester\cite{chattester} and SymPrompt\cite{symprompt}. All methods are asked to generate unit tests for the complex methods in the project-to-test. Evosuite is set to be able to test private methods and utilise Mockito as instructed by the authors\footnote{\url{https://github.com/EvoSuite/evosuite/issues/151}}. gpt-turbo-3.5-0125 is used by all methods using an LLM. The results show that HITS overperforms all baseline methods in the line and branch coverage scores by a percentage ranging from 10 to 20. The ablation study confirms that generating tests for code slices separately helps improve the coverage scores.

Our contributions are summarised as follows:
\begin{itemize}
    \item We first propose generating tests in the unit of code slice with large language models to improve the coverage scores for testing complex methods.
    \item We propose HITS, a practical tool designed specifically for testing complex Java methods using the idea of generating tests slice by slice.
    % \item \kb{we conduct a comprehensive evaluation of test generation approaches on complex focal methods...}
    \item We comprehensively evaluate HITS on complex focal methods, proving its effectiveness in generating high-coverage unit tests.
    % \item \kb{we provide a replication package...}
\end{itemize}

\section{Related Work}
\subsection{Unit Test Generation}
`Unit testing' means testing individual hardware or software units or groups of related units\cite{terms}. The unit tests are developed by developers with regular code\cite{survey}. They help find the errors early in the development process\cite{economic}. Thus, researchers discuss generating unit tests automatically. People hope the automatically generated unit tests can cover all branches and lines of the units to test.

%% TODO: add references
To reach high coverage scores, the researchers have tried search-based software testing (SBST) methods\cite{harman2015achievements,mcminn2011search}, symbolic execution exploring possible execution paths\cite{baldoni2018survey,cadar2008klee,cha2012unleashing,chipounov2012s2e}, direct test generation with deep neural networks\cite{athenatest,hu2023identify,rao2023cat}, etc. For SBST studies, the researchers have developed multiple searching strategies, e.g., the evolution algorithm for Evosuite\cite{evosuite} and coverage-directed randomised test generation for Randoop\cite{randoop}. However, for complex methods, the search space is too broad to explore\cite{mcminn2011search,tang2024chatgptvssbst}. Symbolic execution methods, e.g., Dart\cite{dart}, try to examine feasible execution paths. However, due to the path explosion, methods based on symbolic execution could be more challenging to scalable\cite{cadar2008klee,xie2009fitness}. Deep learning models are tried to generate tests directly given the unit method to test\cite{athenatest,alagarsamy2023a3test}. The methods using deep learning models are scalable due to the end-to-end feature, and they can generate more varied input compared with SBST since they learn from human-written tests from multiple data sources. However, the code generated by the deep learning model is not guaranteed to be compliable and executable\cite{yang2022survey}.

Large language models (LLMs) have shown their potential to generate high-coverage unit tests\cite{wang2024software,schafer2023empirical}. Previous works for code generation have shown that they can generate executable code following human instructions\cite{chen2021evaluating,liu2020deep}. A few works have already tried applying LLMs for unit test generation. ChatTester\cite{chattester} shows that the LLM can outperform the SBST methods. ChatUniTest\cite{chatunitest} provides a practical solution and a published tool for applying ChatGPT to Java unit test generation. Codamosa\cite{codamosa} calls the LLM when the traditional SBST tools for Python do not work. SymPrompt\cite{symprompt} is inspired by symbolic execution and uses execution paths as scaffolds for the LLM. Our work focuses on further leveraging the potential of the LLMs to generate unit tests for the complex Java methods, for which previous methods achieve low coverage scores.

\subsection{Program Decomposition}
`Program Decomposition' literally decomposes the program into multiple slices following specific criteria. For example, Hamilton and Danicic\cite{dep} decompose a program based on the result of community discovery using the program's backwards-slicing graph. Program Decomposition helps narrow down the scope of the analysis and simplify the problem. Yong et al.\cite{slice_bug} decompose a buggy program and build a semantic representation for each slice separately to mitigate the complexity of matching bug reports and the source code. Our work incorporates program decomposition to simplify generating unit tests for a complex method to create tests for several simple code slices.

% `Program Slicing` is the computation of the set of relevant statements. For example, to construct the backward-slicing graph mentioned above, each statement's transitive control and data dependencies are calculated, i.e., the backward slicing. Program Slicing identifies the statements worth inspection, making the analysis time-saving and accurate. Our work incorporates program slicing to determine the conditions to reach specific lines of code.
% \kb{A distinction should be made between the products(outputs) of `decomposition' and `slicing', both being called `slice' can cause confusion}

\section{Methodology}

\begin{figure*}[htbp]
    \centering
    \includegraphics[scale=0.6]{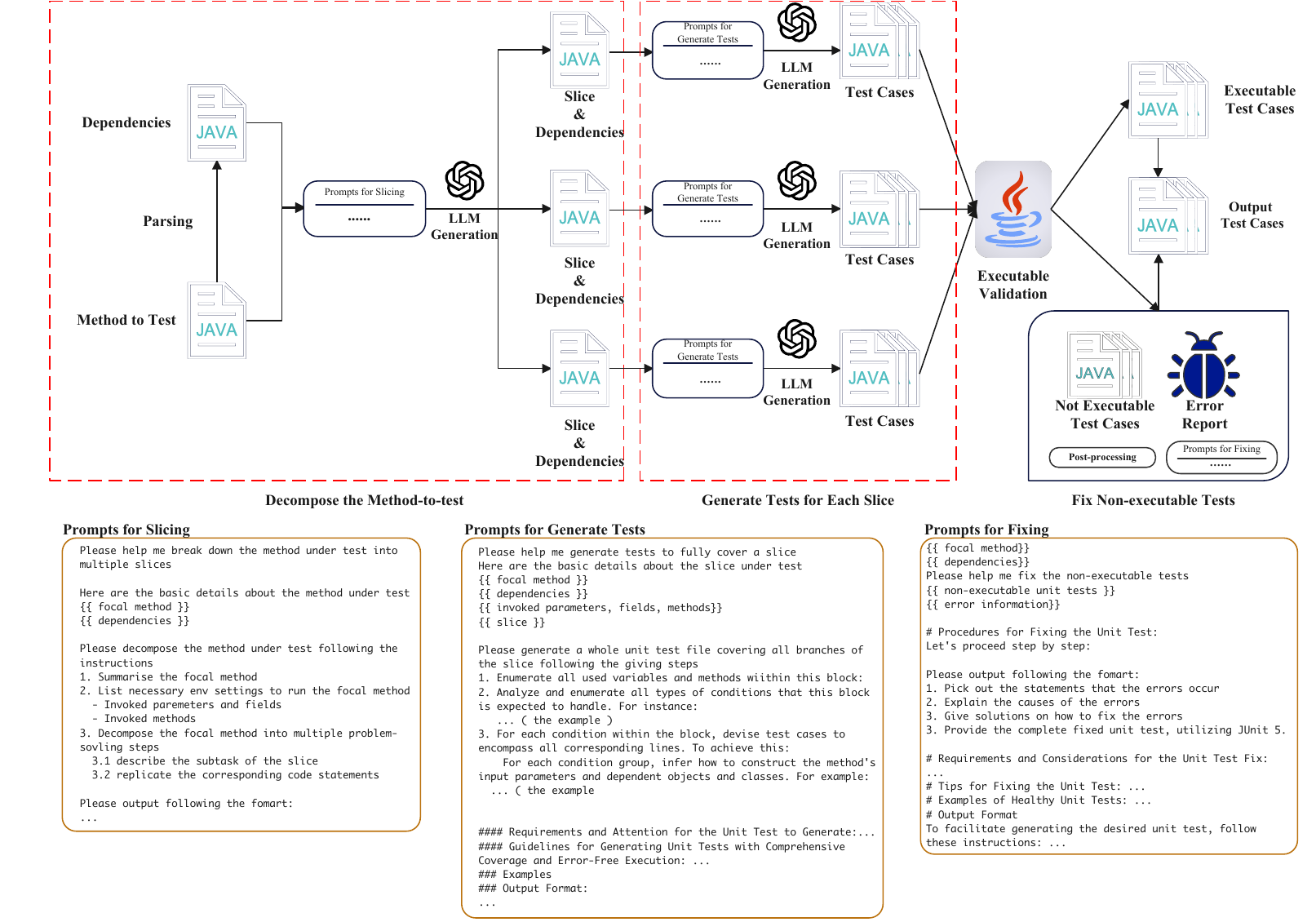}
    \caption{Method Overview}
    \label{fig:method}
\end{figure*}
\subsection{Overview}

HITS generates high-coverage test suites through a two-step process. First, it decomposes the focal method into slices and creates unit tests for each slice, designed to cover all lines and branches. These unit tests collectively form the initial test suite. We then execute all tests in the test suite and divide them into the executable and the non-executable. For the non-executable, HITS includes a fixer to repair them. We illustrate our method in Figure \ref{fig:method}.

In the following sections, we will sequentially provide detailed introductions about HITS

\subsection{Decompose Method into Slices}
\label{sec:method_decompose}
Generating tests for a slice alleviates the difficulty of enumerating input conditions to reach high coverage scores. A slice compresses part of lines of code in the focal method, i.e., a code block. We decompose the method body for a complicated focal method so that each slice $i$ has $n_i$ input condition combinations where $n_i > 1$. If we generate test cases slice by slice, for slice $i$, the large language model needs to infer only one input combination for each input combination of $i$. So, the total combinations to inference is $\sum n_i$. Otherwise, the total number is $\prod n_i$, far more significant than $\sum n_i$. Thus, generating test cases slice by slice makes it easier for the LLM to get high coverage scores. 

% \kb{this para is hard to understand, add a figure to illustrate this idea can be better}

HITS asks the large language models to generate a unit test class to cover a slice of the focal method to cover all lines and branches. Initially, our method parses the focal method via static analysis to collect context for the large language model to understand the usage and structure of the focal method. Subsequently, our method applies Chain-of-Thought to instruct the large language model to decompose the focal method into slices. 

\textit{Context information retrieval.} HITS retrieves context information of the focal method to understand the focal method and build valid test inputs. Like ChatUniTest, we collect the dependent classes' fields and method declarations \cite{chatunitest}. Their bodies are also collected for the methods invoked by the focal method. Above is the information collected by ChatUniTest, and we have adopted it all. Besides, we adopt the Javadocs for the methods invoked if they exist because they give good descriptions for their inputs and outputs in real-world scenarios, helping the LLM to understand the information naturally. The context information and the source code of the focal method serve as the prefix of all prompts for the LLM in HITS.

% \kb{this para is not related to the name of section~\ref{sec:method_decompose}}

\textit{Method Decomposition via Prompting for the LLM.} HITS applies Chain \textit{of thought} to decompose the focal method into slices according to steps for solving a problem, and each slice corresponds to one step.  A method is a solution to a problem that takes several steps. Slicing by steps for solving problems is more natural and human-like, allowing the LLM to understand each slice's functionality and behaviour and use scenarios. We let the LLM decompose the focal method because the LLM has learned about planning, which is difficult to describe in formal languages. To help the LLM understand the decomposition task, we design the \textit{Chain-of-thought} strategy, leading the LLM to work step by step. First, the LLM is asked to summarise the focal method. Second, to understand the method body in detail, we apply the recitation technique, asking the LLM to recite the meaning and usage of all invoked non-local variables and methods. Based on the previous two steps of analysis, the LLM is required to decompose the focal method into slices. We use one-shot prompting here, providing a hand-craft example illustrating how to decompose a solution into several steps. For the simplicity of further analysis and generation, the LLM is instructed to output in JSON format. For each slice, the LLM generates a description in natural language and recites the corresponding original code segment.

% \kb{the name of section~\ref{sec:method_decompose} is the same with this para. the structure of this section~\ref{sec:method_decompose} is confusing }

\subsection{Generate Tests for a Slice}

HITS applies \textit{Chain-of-thought} to instruct the LLM to generate a unit test class for each slice found previously. The plan is designed as follows. Initially, the LLM needs to recite the inputs of the target slice's corresponding code block and the methods invoked by the block. Subsequently, the LLM is instructed to list all possible scenarios of this block. Each scenario corresponds to a feasible input condition combination. The scenario is described in natural language as an 'abstract expression' for the input condition combination, which is human-like and good for the LLM to understand. After that, the LLM inferences how to set the overall execution environment so that when the execution reaches the slice, a specific scenario can be met. Finally, with all the information, the LLM generates a unit test class for the slice required to cover all its lines and branches. We provide examples to help the LLM understand `how to analyse each slice's scenarios' and `how to infer execution environment.' These examples are crafted manually. Besides, we provide tips on generating executable test code considering the features of Java and JUnit 5 test framework, e.g., how to handle private variables, organise a @Test method, etc. Besides that, we present a framework for a correct JUnit 5 test class.

\textit{Fixing Broken Tests via LLMs}. We employ Self-Debug, i.e., let the large language model rectify the broken tests. We adapt Self-Debug by considering the specific characteristics of Java Junit 5 unit tests. The central technique is \textit{Chain-of-thought}. The LLM is provided with the broken test and the error report from the JVM. Besides, it receives a plan to fix the failed test step-by-step. The plan is: 1) to summarise the cause of the error, 2) to describe how to alleviate the cause in natural language, and 3) to generate the fixed version of the test. In addition to the above content, we provide common causes of compilation/execution failures as few-shot examples. We give each cause its description and solution in natural language and an actual example illustrating the fix. These examples are derived from our extensive observations. Though the examples do not encompass all causes of failure, they are still valuable and helpful. The same broken test fixing technique is applied for the patching tests introduced later.

\textit{Instructions and Examples Mitigating Errors from Generating Tests for a Slice}. Since generating tests for a code slice is not common in real-world unit test development, the LLM relies heavily on the instructions and few-shot examples to adapt to slice-coverage-oriented unit test generation. We find that HITS generates tests with complex input construction, making mistakes in Java grammar and common third-party libraries. We summarise the common mistakes into categories and search the Internet for solutions. Then, the solutions are validated and organised as instructions and examples provided to the LLM in the prompt. The categories are: 

\begin{itemize}
    \item the basic structure of an executable JUnit 5 test suite
    \item the usage of reflection to handle private/protected Java code elements
    \item the strategy handling focal methods related to inner classes.
    \item the usage of the Mockito library
\end{itemize}

Finally, The executable and successfully fixed tests are collected as the test suite generated by HITS.

\subsection{Post-Process of LLM's Generation}

We present the techniques applied by HITS of post-processing the text generated by the LLM, which shows the effectiveness and can be generalised to other works applying an LLM.

\textit{Format}. The generation of the LLM can be divided into 1) chain-of-thought analysis and 2) output for extraction, e.g., the description of a slice and the unit test code. We instruct the LLM to generate the chain-of-thought analysis in the Markdown format since it provides an unambiguous hierarchical structure. The LLM must wrap the extraction output in JSON format. Thus, we can easily identify the content needed from plain text with a regex matcher and a JSON decoder. The LLM makes very few mistakes in these two formats, making it very convenient to post-process the text.

\textit{Test Case Identification and Isolation}. We identify and isolate the independent test cases from a whole test file. One test file containing several independent test cases is a feature for Java Junit-style tests. Isolating the test cases helps lessen the workload for further processes. The identification is done by pattern-matching. The isolation is performed by code transformation.

\textit{Rule-based Fixing} Following ChatUniTest \cite{chatunitest}, we apply three rule-based fixing methods for the test cases that failed to compile or execute. They are 1) brackets balancing via pattern matching, 2) removing unnecessary validation since this work focuses on reaching high coverage, and 3) importing commonly used packages.

\section{Experiments}

\subsection{Experiment Setup}
In this section, we first introduce the construction and statistics of the evaluation dataset. Afterwards, we introduce the configuration for generating tests and baselines.

\textbf{Dataset.}
The dataset is constructed following ChatUniTest\cite{chatunitest} and EvoSuite\cite{evosuite}. We crawl ten projects from the Internet and extract their complex methods to form the dataset. The statistics are posted in Table \ref{dataset statistics}. Following \textit{ChatUniTest}, we split the collected projects into two groups. The top 6 projects were created earlier than the cutoff date of the training data of gpt-turbo-3.5. The others were created later than the cutoff date, meaning the LLM has learned nothing about the bottom four projects.

The projects are selected following these standards. First, the project or its domain should be used in Evosuite or ChatUniTest. Second, the project should contain 1 to 30 complex methods. Third, when selecting projects in the domain adapted by ChatUniTest, we should prioritise the projects with high stars on GitHub. The three standards make our projects 1) high quality, 2) reproducible, and 3) reach a balance between variety and budget. We discard the large projects that would cost too many tokens due to our limited budget.

\begin{table*}[htbp]
\caption{Dataset Statistics}
\centering
\begin{tabular}{lllllccc}
\toprule
Project                  & Abbr.~ & Domain                & Version        & Learned? & \#MUTs & Avg. Line    & Avg. PPL  \\ 
\midrule
Commons-CLI              & CLI    & Cmd-line Interface    & 1.5.0          & Y        & 6      & 32.83        & 12.00     \\
Commons-CSV              & CSV    & Data processing       & 1.10.0         & Y        & 6      & 47.67        & 15.67     \\
Gson                     & GSO    & Serialization         & 2.10.1         & Y        & 21     & 51.14        & 20.29     \\
Commons-codec            & COD    & Encoding              & 3a6873e        & Y        & 19     & 73.00        & 21.58     \\
Commons-collections4     & COL    & Utility               & 4.5.0-M1       & Y        & 14     & 28.36        & 14.00     \\
JDom2                    & JDO    & Text Processing (XML) & 2.0.6          & Y        & 21     & 36.71        & 14.19     \\ 
\midrule
Datafaker                & DAT    & Data Generation       & 1.9.0          & N        & 8      & 28.13        & 11.62     \\
Event-ruler              & RUL    & Event Engine          & 1.4.0          & N        & 16     & 75.94        & 24.62     \\
windward                 & WIN    & Microservices         & 1.5.1-SNAPSHOT & N        & 2      & 34.00        & 13.00     \\
batch-processing-gateway & BPG    & Cloud Computing       & 1.1            & N        & 7      & 48.57        & 13.71     \\      
\bottomrule
\end{tabular}
\label{dataset statistics}
\end{table*}
\textbf{Generation Configuration.} We use OpenAI's gpt-turbo-3.5-0125 as the large language model in this work since this model strikes a good balance between the price and the performance. We also apply gpt-turbo-3.5-0125 to all baselines by calling a paid API. We do not truncate the input prompts since gpt-turbo-3.5-0125 can handle long inputs. For reproduction, we use the greedy generation first. If the LLM's responses violate the format provided in section 3.4, we raise the top-p slowly and gradually. We let the LLM generate one unit test file for each slice we find. For the fixing, we allow the LLM to fix broken tests in most ten rounds.

\textbf{Baselines.} We use the state-of-the-art unit test generation methods using an LLM and Evosuite\cite{evosuite} as our baseline. The LLM based baselines are: 
\begin{itemize}
    \item \textbf{ChatUnitTest}: It generates unit tests with an LLM using the focal method and the context extracted by rules as inputs. It uses error reports from JVM as input for the tests that failed to execute to instruct the LLM to fix these tests. 
    \item \textbf{ChatTester}: It is similar to ChatUniTest except for constructing the focal method's context. It constructs the context incrementally as needed.
    \item \textbf{Sympropmt}: Another method generating unit tests with an LLM. SymPrompt divides the program into several control flow paths and instructs the LLM to create unit tests for each path.
\end{itemize}

We implement SymPrompt since we have found no implementations. For ChatUniTest\footnote{\url{https://github.com/ZJU-ACES-ISE/chatunitest-core}} and ChatTester\footnote{\url{https://github.com/ZJU-ACES-ISE/ChatTester}}, we use the open-source implementations. For ChatUniTest and ChatTester, we let them generate 6 unit test files for each method-to-test and fix in most ten rounds. For SymPrompt, we let it generate one unit test file for each path. These settings keep the scales of the test suites at the same level and are good for fair comparison.

\textbf{Research Questions.}
We raise the following research questions to evaluate the effectiveness of our method:

\begin{itemize}
    \item \textbf{RQ1: (Coverage Score Comparison)} Does HITS outperform the baselines on the coverage scores when testing complex methods? Do LLM-based methods all outperform SBST methods when testing complex methods?
    \item \textbf{RQ2: (Execution Correctness Check)} How many test cases generated HITS are executable compared with baseline methods using LLMs? What's the relationship between the proportion of executable tests and total coverage scores?
    \item \textbf{RQ3: (Ablation Study)} How does each component of HITS contribute to the final performance?
\end{itemize}

\subsection{RQ1: Coverage Scores Comparison}
\begin{table}
\centering
\caption{Test Case Count}
\begin{tabular}{lccccc} 
\toprule
Proj.                & \makecell[l]{SlicePrompt-\\Test4j} & \makecell[l]{Chat-\\UniTest} & \makecell[l]{Chat-\\Tester} & \makecell[l]{Sym-\\Prompt} & Evosuite \\ 
\midrule
CLI          & 64                & 168         & 68         & 1779   & 87    \\
CSV          & 62                & 138         & 66         & 1297   & 29    \\
GSO          & 225               & 590         & 244        & 1226   & 207    \\
COD        & 167               & 326         & 230        & 366     & 325   \\
COL & 102               & 308         & 136        & 306    & 159    \\ 
JDO                & 205               & 386         & 268        & 1183   & 127    \\
\midrule
DAT            & 54                & 140         & 100        & 331    & 85    \\
RUL          & 134               & 364         & 198        & 361   & 210     \\
WIN             & 24                & 64          & 16         & 104   & 6     \\ 
BPG                  & 68                & 118         & 76         & 800  & N/A      \\
\bottomrule
\end{tabular}
\label{tab:test case count}
\end{table}

\begin{table*}[htbp]
\centering
\caption{Line Coverage Scores on Complex Methods}
\begin{tabular}{lccccc} 
\toprule
Project Abbr.     & HITS & ChatUniTest      & ChatTester & SymPrompt & Evosuite         \\ 
\midrule
CLI              & \textbf{77.83\%}  & 51.17\%          & 50.33\%    & 47.00\%      & \underline{53.50\%}   \\
CSV              & \textbf{57.67\%}  & 19.67\%          & 16.67\%    & 30.50\%    & \underline{36.20\%}   \\
GSO                    & \textbf{58.61\%}  & \underline{44.43\%}  & 21.05\%    & 25.05\%   & 39.00\%             \\
COD           & \underline{67.84\%}   & 26.26\%          & 30.05\%    & 43.00\%      & \textbf{79.30\%}  \\
COL      & \underline{43.71\%}   & \textbf{46.21\%} & 38.93\%    & 24.93\%   & 31.70\%           \\
JDO                    & \textbf{43.38\%}  & 29.52\%          & 23.52\%       & 26.24\%   & \underline{36.10\%}   \\
\midrule
DAT               & \underline{54.75\%}   & 47.25\%          & 24.50\%     & 17.25\%   & \textbf{61.80\%}  \\
RUL            & \textbf{24.13\%}  & 5.56\%           & 2.00\%        & 3.94\%    & \underline{14.30\%}   \\
WIN                & \textbf{52.00\%}     & 0.00\%              & 0.00\%        & 1.50\%     & 0.00\%              \\
BPG & \textbf{71.00\%}     & \underline{54.71\%}  & 0.00\%        & 43.71\%   & N/A              \\ 
\midrule
Avg.                     & \textbf{55.09\%}  & 32.48\%         & 20.71\%    & 26.32\%  & \underline{39.10\%}           \\
\bottomrule
\end{tabular}
\label{tab:line cov}
\end{table*}

\begin{table*}[htbp]
\centering
\caption{Branch Coverage Scores on Complex Methods}
\begin{tabular}{lccccc} 
\toprule
Project Abbr.            & HITS & ChatUniTest     & ChatTester & SymPrompt & Evosuite          \\ 
\midrule
CLI              & \textbf{73.00\%}   & 45.83\%         & 47.17\%    & 46.33\%        & \underline{52.50\%}    \\
CSV              & \textbf{52.33\%}  & 15.00\%          & 15.00\%     & 29.00\%         & \underline{36.50\%}   \\
GSO                     & \textbf{52.90\%}  & \underline{41.24\%} & 20.19\%    & 23.67\%        & 37.71\%           \\
COD            & \underline{58.63\%}   & 20.47\%         & 25.00\%    & 38.63\%        & \textbf{79.32\%}  \\
COL      & \underline{33.64\%}   & \textbf{37.50\%} & 31.26\%    & 21.50\%         & 29.50\%            \\
JDO                    & \textbf{39.90\%}  & 26.23\%         & 20.50\%       & 25.95\%        & 35.33\%           \\
\midrule
DAT               & \underline{45.13\%}   & 39.00\%          & 20.50\%     & 20.63\%        & \textbf{64.88\%}  \\
RUL             & \textbf{17.63\%}  & 4.13\%          & 0.94\%     & 1.75\%         & \underline{10.36\%}   \\
WIN                & \textbf{49.00\%}   & 0.00\%             & 0.00\%        & \underline{1.50\%}  & 0.00\%               \\
BPG & \textbf{59.00\%}     & 41.29\%         & 0.00\%        & \underline{42.00\%} & N/A               \\
\midrule
Avg.                     & \textbf{48.12\%} & 27.07\%        & 18.20\%     & 25.10\%       & \underline{38.46\%}   \\
\bottomrule
\end{tabular}
\label{tab:branch cov}
\end{table*}

We report, compare and analyse the coverage scores testing complex Java methods. This is the fundamental experiment of this paper, which directly assesses the effectiveness of HITS. We separately report the line and branch coverage scores for better reading experiences. Both scores are calculated with Jacoco \footnote{\url{https://github.com/jacoco/jacoco}}, an open-source tool measuring coverage scores for Java.

We report the line coverage scores in Table \ref{tab:line cov} and Table \ref{tab:branch cov}. Each row is for a project, and each column is for a method. The best result on a project-to-test is bold, and the second-best result is underlined. Notably, we don't report Evosuite's results on the project `batch-processing-gateway' since it is not executable on Java 8, but Evosuite relies on Java 8.

According to Table \ref{tab:line cov} and Table \ref{tab:branch cov}, our method outperforms all baselines relying on LLMs and the SBST method, Evosuite, comparing the average scores. To dive deep into the results, our method gives the best results on seven project-to-test and the second-best results on the remaining 3. 

Compared with LLM-based baseline methods, our method is better than all on all projects except `Commons-collections'. We analyse that the `Commons-collections' is more broadly used than the other projects, and the LLM has learned enough about the project from its training data. Thus, `recalling' various test cases from its memory of the training data is more natural and complete than the instruction we construct to generate high-coverage tests. 

We have noticed that our method outperforms the SBST method and Evosuite in the overall scores but has performance gaps on specific projects, e.g., `Datafaker'. Evosuite is better than all LLM-based test generation methods on these specific projects. Besides, considering the average scores, the second best method is Evosuite rather than the LLM-based methods. These results imply that although when testing on all methods in a project, the LLM-based methods have an advantage over the SBST methods, there still exist challenges to testing complex methods. Thus, our method is a good start to further improve LLM-based test generation methods by focusing on testing complex methods.

Additionally, we report the number of test cases generated by each LLM-based Test Generation Method to prove that HITS does not use sampling many test cases to gain advantages. The report is in Table \ref{tab:test case count}, showing that HITS does not prominently sample more test cases than the baselines. We do not compare the number of test cases with Evosuite since the stopping criteria for Evosuite is its search reaches a convergence. However, the LLM-based methods are constrained by generation rounds. Under two different criteria of stopping generation, the comparison between Evosuite and LLM-based methods is unfair. We list the figures of Evosuite only for references.

\begin{myresultbox}[Answer to RQ1:]
HITS outperforms all the baselines on the coverage scores. When testing complex methods, the LLM-based baselines are inferior to Evosuite, making HITS's improvement valuable for further development on LLM-based test generation methods.
\end{myresultbox}

\subsection{RQ2: Execution Correctness Check}
We compare the execution correctness among the LLM-based test generation methods. The unit tests are generated via the LLM's auto-regressive decoding without explicit constraints to guarantee the code's compilation \& runtime correctness. Thus, the tests generated by the LLM are not guaranteed to be executable, and increasing the portion of executable tests is a target for LLM-based test generation methods, including our work. 

We count and calculate the portion of successfully executed test cases for each method, i.e. the \textbf{pass rate}. Then, we compare the distribution of the two types of error: compilation error and runtime error. 

The proportions of executable tests are reported in Table \ref{tab:pass rate}. The table shows that our method, HITS, has advantages over the LLM-based baselines of higher pass rates on average. When taking Table \ref{tab:branch cov} and Table \ref{tab:line cov} into consideration, we have the findings that 1) low pass rates lead to low coverage scores and 2) high pass rates can not guarantee high coverage scores. Designing a single test case to cover all branches and lines is challenging when testing complex methods. Low pass rates mean insufficient test cases, resulting in low coverage scores. With a limited time budget for unit test generation tools, it is necessary to increase the pass rate. However, high pass rates are not equivalent to high coverage scores. Taking ChatUniTest on Commons-CLI as an example, though Commons-CLI reaches the second-best pass rate scores, its coverage scores are still lower than Evosuite's. Several executable tests likely cover the same branches and lines of the focal method. Thus, guidance for LLM in generating unit tests for specific purposes is necessary to avoid similar test cases.

\begin{myresultbox}[Answer to RQ2:]
HITS outperforms all the baselines on the pass rate, considering the average scores. Low pass rates lead to low coverage scores, while high pass rates do not guarantee high ones. 
\end{myresultbox}

\begin{table*}
\centering
\caption{Pass Rate Comparison on Complex Methods}
\begin{tabular}{lcccc} 
\toprule
Project Abbr.                & HITS & ChatUniTest     & ChatTester & SymPrompt         \\ 
\midrule
CLI              & \textbf{100.00\%} & \underline{85.71\%} & 41.18\%    & 33.33\%           \\
CSV              & \textbf{82.04\%}  & \underline{34.78\%} & 9.09\%     & 16.11\%           \\
GSO                     & \textbf{64.89\%}  & \underline{47.37\%} & 14.75\%    & 11.66\%           \\
COD            & \textbf{82.04\%}  & \underline{57.67\%} & 23.48\%    & 41.80\%           \\
COL & 55.88\%           & \underline{59.74\%} & 47.06\%    & \textbf{66.67\%}  \\
JDO                   & \textbf{60.98\%}  & \underline{34.29\%} & 13.43\%    & 1.52\%            \\ 
\midrule
DAT               & \textbf{59.26\%}  & \underline{34.29\%} & 8.00\%     & 5.44\%            \\
RUL              & \textbf{34.33\%}  & \underline{21.98\%} & 4.04\%     & 13.30\%           \\
WIN                 & \textbf{95.83\%}  & 0.00\%          & 0.00\%     & \underline{2.88\%}    \\
BPG & \textbf{55.88\%}  & \underline{42.37\%} & 0.00\%     & 17.25\%           \\ 
\midrule
Avg.                     & \textbf{69.11\%}  & \underline{41.82\%} & 16.10\%    & 21.00\%           \\
\bottomrule
\end{tabular}
\label{tab:pass rate}
\end{table*}

\subsection{RQ3: Ablation Study}
We analyse the components' contribution to improving coverage scores via the ablation study. Compared with ChatUniTest, which directly inspires us with the fundamental workflow, HITS introduces three original techniques to generate high-coverage score unit tests for complex methods. They are 1) the slicing technique, i.e. the workflow for slicing the focal method 2) the prompt engineering (PE), i.e. the instructions and examples to teach the LLM generate as we expect; and 3) the post-processing(PP), including formatting LLM's output, test case extraction and rule-based fixing. We perform the ablation study to prove their contribution. 

We perform the experiments under two configurations. One is `w/o slicing', i.e. removing all components for slicing. We want to know how much contribution is made by the idea of `slicing' only. The other is `w/o slicing \& PE'. We remove the workflow for slicing and the instructions and examples to teach the LLM to generate tests. `w/o slicing \& PE' is equivalent to `ChatUniTest + PP', which should be the comprehensive manifestation of ChatUniTest. Since our PE is explicitly designed for teaching the LLM to learn generating for slices and the post-processing can be applied to any LLM-based method, this configuration can compare the PE's contribution and the contribution of the whole technique suite for slicing.

The result is posted in Table \ref{tab:ablation}. On the average scores, compared with the fundamental work ChatUniTest (32.48\% for line coverage and 27.07\% for branch coverage), all components contribute to the coverage scores. When regarding `ChatUniTest + PP' as the complete form of `ChatUniTest', we find that the workflow of slicing a method contributes more than the instructions and examples. Then, we dive into the analysis for individual projects. For all projects, the slicing workflow + PE improve the coverage scores. However, we also view that the slicing workflow affects the branch coverage project scores for a few projects. We analyse and get the following reasons. First, these projects are in the LLM's training set, so it's easier and more natural for the LLM to recall the tests and ways of using the methods-to-test than building tests from scraps following the slicing workflow. This can be reflected by the consistent improvements in the dataset, which the LLM does not learn. Second, additional information, i.e. the additional prompt content for slicing, makes the input for LLMs grow longer, which is unfriendly to LLMs. The improvements brought by PP can reflect this. Our PP helps shorten the inputs for the LLM compared with ChatUniTest, and the gain is much. Making the LLM learning to generate tests slice by slice in other ways rather than in-context learning remains our future work. As for the drawback of line coverage scores, since the numbers of lines included by branches are different, we present them for a quick look by convention rather than a more profound analysis since a high branch coverage score is not equivalent to a high line coverage score. Our work pays more attention to branch coverages and regards all branches equally. Choosing branches with more lines to cover first remains our future work.
\begin{table*}
\centering
\caption{Ablation Study}
\begin{tabular}{llcccccc} 
\toprule
\multirow{3}{*}{\makecell[l]{Project\\Abbr.}} & \multirow{3}{*}{\makecell[l]{LLM \\Learned?}} & \multicolumn{3}{c}{Line Coverage}                                                                           & \multicolumn{3}{c}{Branch Coverage}                                                                           \\ 
\cline{3-8}
                                &                                                                         & HITS & w/o slicing & \makecell{w/o slicing PE\\ ChatUniTest + PP} & HITS & w/o slicing & \makecell{w/o slicing PE\\ ChatUniTest + PP}  \\ 
\midrule
CLI                             & Y                                                                       & 77.83\%           & 57.67\%     & 58.50\%                                                                   & 73.00\%            & 56.50\%      & 55.17\%                                                                     \\
CSV                             & Y                                                                       & 57.67\%           & 53.66\%     & 42.17\%                                                                   & 52.33\%           & 30.33\%     & 30.33\%                                                                     \\
GSO                             & Y                                                                       & 58.61\%           & 59.48\%     & 57.52\%                                                                   & 52.90\%           & 53.48\%     & 51.19\%                                                                     \\
COD                             & Y                                                                       & 67.84\%           & 65.58\%     & 56.26\%                                                                   & 58.63\%           & 55.26\%     & 50.21\%                                                                     \\
COL                             & Y                                                                       & 43.71\%           & 46.29\%     & 46.85\%                                                                   & 33.64\%           & 37.93\%     & 38.93\%                                                                     \\
JDO                             & Y                                                                       & 43.38\%           & 45.28\%     & 45.80\%                                                                   & 39.90\%           & 42.48\%     & 42.10\%                                                                     \\ 
\midrule
DAT                             & N                                                                       & 54.75\%           & 50.38\%     & 48.50\%                                                                    & 45.13\%           & 42.00\%      & 40.50\%                                                                      \\
RUL                             & N                                                                       & 24.13\%           & 20.63\%     & 20.13\%                                                                   & 17.63\%           & 15.86\%     & 14.69\%                                                                     \\
WiN                             & N                                                                       & 52.00\%            & 58.00\%      & 47.00\%                                                                    & 49.00\%            & 47.00\%      & 46.50\%                                                                      \\
BPG                             & N                                                                       & 71.00\%              & 49.14\%     & 62.86\%                                                                   & 59.00\%              & 45.86\%     & 54.14\%                                                                     \\ 
\midrule
Avg.                            &             /                                                            & 55.09\%          & 50.61\%    & 48.56\%                                                                  & 48.12\%          & 42.97\%    & 42.38\%                                                                    \\
\bottomrule
\end{tabular}
\label{tab:ablation}
\end{table*}

\begin{myresultbox}[Answer to RQ3:]
All components of HITS, including the slicing workflow, the instructions \& examples and the post-processing, contribute to the performance improvements in the average coverage scores. The detailed comparison for specific testing projects shows that HITS can reform how teaching the LLM generates tests slice by slice for further improvements.
\end{myresultbox}
\section{Discussion}
\subsection{Case Study}

\begin{figure*}[htbp]
    \centering
    \includegraphics[scale=0.9]{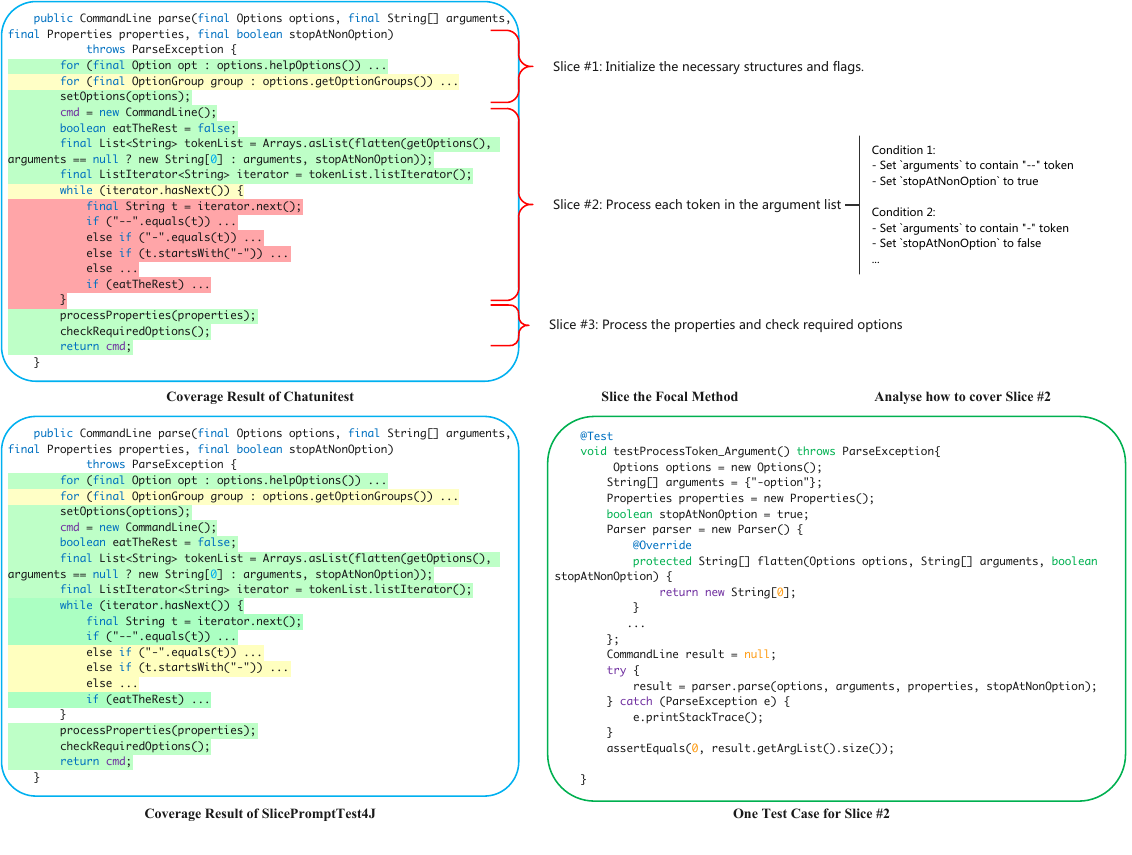}
    \caption{Case study}
    \label{fig:case study}
\end{figure*}

We use a case illustrated in Figure \ref{fig:case study} to show how HITS outperforms the baselines via generating tests for a slice. The example method-to-test is \lstinline|Parser.parse()| of CLI.

First, we show the code coverage status of ChatUniTest's test suite. The green line means all lines \& branches are covered. The yellow line means `partially covered', and the red line means `zero coverage'. The figure shows that ChatUniTest's test suite fails to cover the code block within \lstinline|while (iterator.hasNext())|. ChatUniTest does not generate test cases, which makes \lstinline|tokenList| non-empty.

By focusing on generating tests for a slice, HITS builds test cases that cover the code block missed by ChatUniTest. Initially, HITS decomposes the focal method into slices, as shown in the figure, and the missed block lies in Slice \#2. When generating tests for Slice \#2, HITS concentrates on analysing the conditions in Slice \#2 for high coverage scores. To make \lstinline|tokenList| non-empty and contain various elements to cover the complex branches within the while-loop, HITS inferences multiple conditional groups of the input parameters, \lstinline|arguments| and \lstinline|stopAtNonOption|. We give ONE of the generated test cases for Slice \#2. In the example, a short option recognised as a Java parameter is sent to the focal method, entering the missing code block and covering the corresponding branches. 

The coverage status of HITS is posted beneath ChatUniTest's status. By focusing on Slice \#2 and generating thoughtfully designed tests, the missed code block is partially covered, and the overall coverage scores are improved.
\subsection{Methodology for Optimizing Prompts}

%% TODO: add references
\textit{Prompt Self-refine.} Self-refine \cite{selfrefine} helps the LLM better understand our intention. Self-refine is the technique of asking the LLM to optimise the prompt itself. Our early observations found that the LLM violated our output format requirements (e.g., the JSON format) or returned zero slices, empty code and irrelevant analysis. We then tried to improve our prompt with Self-refine by telling the LLM, `Please help me refine the prompt with more fluent expressions and a clearer structure'. The LLM optimised the prompt. With the optimised prompt, the LLM could generate output following our given format and perform analysis as we instruct. Since our order and instructions are complicated and challenging to understand at first glance, Self-refine helps align the LLM and the human, and the refined prompt can drive the LLM to work following the human's instructions as expected.

\begin{table*}[htbp]
\centering
\caption{Non-executable Test Distribution}
\begin{tabular}{lcccccccc} 
\toprule
\multirow{2}{*}{Proj.} & \multicolumn{2}{c}{HITS} & \multicolumn{2}{c}{ChatUniTest} & \multicolumn{2}{c}{ChatTester} & \multicolumn{2}{c}{SymPrompt}  \\ 
\cline{2-9}
                       & Comp. Error & Run. Error              & Comp. Error & Run. Error        & Comp. Error & Run. Error       & Comp. Error & Run. Error       \\ 
\midrule
CLI                    & 0.00\%      & 0.00\%                  & 100.00\%    & 0.00\%            & 65.00\%     & 35.00\%          & 48.74\%     & 51.26\%          \\
CSV                    & 53.85\%     & 46.15\%                 & 100.00\%    & 0.00\%            & 70.00\%     & 30.00\%          & 62.04\%     & 37.96\%          \\
GSO                    & 43.04\%     & 56.96\%                 & 100.00\%    & 0.00\%            & 92.31\%     & 7.69\%           & 67.68\%     & 32.32\%          \\
COD                    & 33.33\%     & 66.67\%                 & 100.00\%    & 0.00\%            & 65.91\%     & 34.09\%          & 25.35\%     & 74.65\%          \\
COL                    & 77.78\%     & 22.22\%                 & 100.00\%    & 0.00\%            & 97.22\%     & 2.78\%           & 45.10\%     & 54.90\%          \\
JDO                    & 81.25\%     & 18.75\%                 & 100.00\%    & 0.00\%            & 75.00\%     & 25.00\%          & 70.90\%     & 29.10\%          \\
\midrule
DAT                    & 90.91\%     & 9.09\%                  & 100.00\%    & 0.00\%            & 91.30\%     & 8.70\%           & 79.55\%     & 20.45\%          \\
RUL                    & 61.36\%     & 38.64\%                 & 100.00\%    & 0.00\%            & 97.89\%     & 2.11\%           & 63.26\%     & 36.74\%          \\
WIN                    & 0.00\%      & 100.00\%                & 100.00\%    & 0.00\%            & 100.00\%    & 0.00\%           & 87.13\%     & 12.87\%          \\
BPG                    & 60.00\%     & 40.00\%                 & 100.00\%    & 0.00\%            & 100.00\%    & 0.00\%           & 73.41\%     & 26.59\%          \\
\bottomrule
\end{tabular}
\label{tab:error dist}
\end{table*}

\subsection{Non-executable Test Distribution}
To further improve LLM-based test generation tools, we analyse the distribution of the non-executable tests. We report the proportion of the test cases with compile (Comp.) error and runtime (Run.) error among the non-executable tests for each LLM-based test generation method used in this paper. The results are illustrated in Table \ref{tab:error dist}.

The more complex the prompt for an LLM to generate tests, the more chances the non-executable tests have runtime errors. According to Table \ref{tab:error dist}, the proportion of runtime errors within the non-executable tests of HITS and SymPrompt is more significant than that of ChatUniTest and Chattester. The principal difference between the two groups is that HITS and SymPrompt have more requirements for the unit test to generate than ChatUniTest and ChatTester. The LLM does not encounter such requirements in the pre-training phase, so they can only learn via in-context learning. However, to fulfil the criteria in the prompt, the constructions of the focal classes are complicated, resulting in improper test case structures leading to runtime errors.

The analysis inspires us that teaching the LLM following instructions more effectively and efficiently is the key to improving LLM-based test generation tools' performances further. We see it as our future work.

\subsection{Threats to Validity}

The threats to the validity of HITS are the limited dataset and generalisation to other LLMs. Due to our limited budget of the experiment fund, we discard the projects with more than 30 complex methods to strike a balance between the number of total methods to test and the projects to test. For the LLM, we use gpt-3.5-turbo because it also strikes a good balance between performance for learning the new generation paradigm and the price for calling API. The other LLMs are either incapable of understanding the instructions of HITS to generate or are much more expensive than gpt-3.5-turbo. Finding ways to extend the dataset and the tested LLM remains part of our future work.

\section{Conclusion}
We point out that current LLM-based unit test generation tools have low coverage scores when testing complex scores. To solve the problem, we propose decomposing the method-to-test into slices and generating unit tests slice by slice, applying the `divide-and-conquer' algorithm. Then, we propose our implementation, HITS, to generate high-coverage unit tests for complex Java methods using an LLM. We have thoroughly evaluated HITS compared with the state-of-the-art LLM-based test generation tools and Evosuite and proved HITS's effectiveness. Our further ablation studies confirm the contribution brought by the idea of generating tests for a slice.
%%
%% The next two lines define the bibliography style to be used and

\bibliographystyle{ACM-Reference-Format}
\bibliography{main}

\end{document}